\documentclass[12pt]{article}
\usepackage[]{psfig}
\newcommand{\be}{\begin{equation}}
\newcommand{\ee}{\end{equation}}
\newcommand{\ba}{\begin{eqnarray}}
\newcommand{\ea}{\end{eqnarray}}

\begin{document}
%%%%%%%%%%%%%%%%%%%%%%%%%%%%%%%%%%%%%%%%%%%%%%%%%%%%%%%%%%%%%%%%%%%
\title{\bf Comments on the $U(2)$ Noncommutative Instanton}
\author{D.H.~Correa$^a$\thanks{ANPCyT} \,,
G.~Lozano$^b$\thanks{Associated with CONICET} \, , \\
E.F.~Moreno$^a\dagger$ and
F.A.Schaposnik$^a$\thanks{Associated with CICPBA}\\
{\normalsize\it $^a$Departamento de F\'\i sica, Universidad Nacional
de La Plata}\\
{\normalsize\it C.C. 67, 1900 La Plata, Argentina}
\\
{\normalsize\it $^b$Departamento de F\'\i sica, FCEyN, Universidad de
Buenos Aires}\\
{\normalsize\it Pab.1, Ciudad Universitaria,
Buenos Aires,Argentina}
}

\date{\hfill}
\maketitle

%===================================================================
\begin{abstract}
We  discuss  the 't Hoof ansatz for instanton solutions in
noncommutative $U(2)$ Yang-Mills theory. We show that the
extension of the ansatz leading to singular solutions in the
commutative case, yields to non self-dual (or self-antidual)
configurations in noncommutative space-time. A proposal leading to
selfdual solutions with $Q=1$ topological charge (the equivalent
of the regular BPST ansatz) can be engineered, but in that case
the gauge field and the curvature are not Hermitian (although the
resulting Lagrangian is real).
\end{abstract}
\date{}

%

%\maketitle
%\pacs{PACS numbers:\ \  11.15.Kc,  14.80.Hv, 11.10.Lm}

%%
%\newpage

%===================================================================

\section{Introduction}
After the connection between noncommutative quantum field theory
and string theory was discovered \cite{CDS}-\cite{DH}, instantons
in gauge theories, originally introduced in \cite{NS} for
noncommutative $R^4$ space,  were seen to play a central r\^ole in
the quantization of strings ending in D-branes in the presence of
a $B$-field \cite{SW}. Later o, they were the object of many
investigations \cite{GN}-\cite{AS}.

Although one can envisageto construct instantons for $U(N)$ gauge
group for arbitrary $N$, most of the results reported correspond
to the case of $U(1)$, for which, in contrast with what happens in
ordinary space, there also exist non-trivial multi-instantons. The
explicit solutions were constructed mostly by applying the ADHM
recipe.

Concerning solutions for $N \geq 2$, appart from discussions on
the ADHM construction \cite{F}-\cite{Furuu},\cite{K},
 the possibility of extending the so called 't Hooft
ansatz for multi-instanton solutions to the noncommutative case was already
suggested in \cite{NS}. However, we will show that
the ansatz proposed in that article for
the case of $U(2)$
leads
to a configuration which is not self-dual (or self-antidual) and hence does
not correspond to a bound of the action.

It is the purpose of this note to carefully test the 't Hooft
ansatz for $U(2)$ noncommutative gauge theory, showing that the naive
extension of the ordinary ansatz leads to a non self-dual
(or non self-antidual) configuration which does not
extremize the action.
The problem  cannot be solved by projecting out an appropriate
state from the Fock space, as it can be done in the ADHM approach
(see \cite{Furuu}
and references therein).
Interestingly enough, although the resulting topological
charge is $Q = 0$, the configuration does
coincide with the ordinary $Q = 1$ instanton
solution in the singular gauge
when the non-commutative parameters
$\theta_{\mu\nu}$ are put to zero. This shows
that the $\theta_{\mu\nu} \to 0$ limit is not smooth.
We shall then analyse an alternative -Belavin-Polyakov-Schwarz-Tyupkin (BPST)
 like- ansatz for $U(2)$
selfdual (self-antidual) solutions which requires a different
internal $U(2)$ structure for the gauge field.  However,
hermiticity of the gauge fields and curvature is lost, although
the resulting instanton Lagrangian is real. Finally, we discuss
possible issues which, starting from BPST like ansatz, may lead to
hermitian gauge field configurations corresponding to
noncommutative instanton.

\section{'t Hooft Ansatz in commutative space}

The first order instanton equations for (ordinary) $SU(2)$ Yang-Mills
theory in 4 dimensional Euclidean space are
\begin{equation}
F_{\mu \nu} = \pm \tilde F_{\mu \nu}
\label{du}
\end{equation}
with
\begin{equation}
F_{\mu\nu} = \partial_\mu A_\nu - \partial_\nu A_\nu
+ i[A_\mu,A_\nu]
\label{fmn}
\end{equation}
and
\begin{equation}
\tilde F_{\mu\nu} = \frac{1}{2} \varepsilon_{\mu\nu\alpha\beta}
F_{\alpha \beta}
\label{eq}
\end{equation}
Here $A_\mu$ are Hermitian gauge fields taking
 values in the Lie algebra of $SU(2)$,
$A_\mu = A_\mu^a \sigma^a/2$, with $\sigma^a$ the Pauli
matrices.

The self-dual ($+$ sign) and self-antidual ($-$ sign) solutions
to equations (\ref{du}) correspond to positive and negative
topological charge $Q$,
\begin{equation}
Q = \frac{1}{16 \pi^2} \!\int \!d^4 x \,{\rm tr} F_{\mu\nu}\tilde F_{\mu \nu}
\label{Q}
\end{equation}

The well-honnored 't Hooft ansatz \cite{tH}-\cite{Jackiw} for
a \underline{self-dual} multi-instanton solution is
\begin{equation}
A_\mu(x) = \bar \Sigma_{\mu \nu}  j_\nu
\label{tH}
\end{equation}
\begin{equation}
j_\nu= \phi^{-1}(x)\partial_\nu  \phi(x)
\label{tHj}
\end{equation}
Here
\begin{eqnarray}
\bar \Sigma_{\alpha\beta} &=&
\frac{1}{2}\sigma^a \bar \eta_{a\alpha\beta}
\label{defi}
\end{eqnarray}
where $\bar \eta_{a\mu\nu}$  is the 't Hooft tensor,
\begin{eqnarray}
\bar \eta_{a\mu\nu} = \varepsilon _{a\mu\nu} \; &,& {\rm if~~} \mu,\nu=1,2,3
\nonumber\\
\bar \eta_{a\mu4} = - \bar \eta_{a 4 \mu } =  -\delta_{a\mu}  \;&,& \;\;\;
\;\;\; \bar \eta_{a44}= 0
\label{bareta}
\end{eqnarray}
It is important to stress that $\bar \Sigma_{\alpha\beta}$ is
anti-selfdual,
\begin{equation}
\bar \Sigma_{\alpha\beta} =- \tilde{\bar \Sigma}_{\alpha\beta}
\end{equation}
Inserting ansatz (\ref{tH}) in eq.(\ref{fmn}) one finds that the curvature
takes the form
\be
F_{\mu \nu}^a = \bar \eta_{a \mu \beta}  v_{\beta \nu}[\phi]
 -\bar \eta_{a \nu \beta} v_{\beta \mu}[\phi]  -
 \frac{1}{2} \bar \eta_{a \mu\nu } v_{\beta \beta}[\phi]
  -  \frac{1}{2}\bar \eta_{a\mu\nu} a[\phi]
\label{vefe}
\ee
where
\be
v_{\mu \sigma} = -\frac{2}{\phi^2} \partial_\mu \phi \partial_\sigma \phi
 + \frac{1}{\phi}\partial_\mu\partial_\sigma \phi
 \label{vv}
 \ee
 \be
 a[\phi] = \frac{1}{\phi} \nabla^2 \phi
 \label{aax}
 \ee
It is easy to prove that $F_{\mu\nu}^a$ as given in (\ref{vefe}) can be written
in the form
\be
F_{\mu\nu}^a = \tilde F_{\mu\nu}^a - \bar \eta_{a\mu\nu} a[\phi]
\label{dualis}
\ee
so that, in order to have selfduality, one should impose
$a[\phi] = 0$  or
\begin{equation}
\frac{1}{\phi} \nabla^2 \phi = 0
\label{fi}
\end{equation}
A general  solution to this equation is
\begin{equation}
\phi(x) = 1 + \sum_{i=1}^N \frac{\lambda_i^2}{\left(x^\mu - a_i^\mu\right)^2}
\label{sing}
\end{equation}
Although this solution has singularities at $x = a_i$
($\nabla^2 \phi =
\sum
\delta (x - a_i)
$),
eq.(\ref{fi})
is satisfied everywhere.

This ansatz leads to a singular self-dual gauge field.
It is instructive to write it in the simplest $Q=1$  case
(with $a_1 = 0$)
\begin{equation}
A^{sing}_\alpha(x) = -2 \lambda^2 \bar \Sigma_{\alpha\beta}
\frac{x_\beta}{x^2(x^2 + \lambda^2)}
\label{Asin}
\end{equation}
Now, the singularity can be removed by an appropriate (singular)
gauge transformation leading to
\begin{equation}
A^{reg}_\alpha(x) = -2 \Sigma_{\alpha\beta}
\frac{x_\beta}{x^2 + \lambda^2}
\label{Areg}
\end{equation}
which coincides with the BPST instanton solution.
It is important to note that the self-antidual $\bar \Sigma_{\alpha\beta}$
tensor appearing in the singular ansatz has been traded for a self-dual
tensor
$\Sigma_{\alpha\beta}$,
\begin{eqnarray}
\Sigma_{\alpha\beta} &=&
\frac{1}{2}\sigma^a  \eta_{a\alpha\beta}
\label{defid}
\end{eqnarray}
\begin{eqnarray}
\eta_{a\mu\nu} = \varepsilon _{a\mu\nu} \; &,& {\rm if~~} \mu,\nu=1,2,3
\nonumber\\
 \eta_{a\mu4} = -  \eta_{a 4 \mu } =  \delta_{a\mu}
 \;&,& \;\;\;
\;\;\; \eta_{a44}= 0
\label{eta}
\end{eqnarray}
The same procedure can be applied  to get regular
multi-instanton solutions with arbitrary topological charge $Q$
\cite{GR}.

%%%%%%%%%%%%%%%%%%%%%%%%%%%%%%%%%%%%%%%%%%%%%%%
%%%%%%%%%%%%%%%%%%%%%%%%%%%%%%%%%%%%%%%%%%%%%%%
%%%%%%%%%%%%%%%%%%%%%%%%%%%%%%%%%%%%%%%%%%%%%%%
%%%%%%%%%%%%%%%%%%%%%%%%%%%%%%%%%%%%%%%%%%%%%%%
%%%%%%%%%%%%%%%%%%%%%%%%%%%%%%%%%%%%%%%%%%%%%%%
%%%%%%%%%%%%%%%%%%%%%%%%%%%%%%%%%%%%%%%%%%%%%%%
%%%%%%%%%%%%%%%%%%%%%%%%%%%%%%%%%%%%%%%%%%%%%%%

\section{'t Hooft Ansatz in noncommutative space}
Let us start by defining the Moyal $*$ product of two functions
$f(x)$ and $g(x)$,
\begin{equation}
\left.(f*g)(x)  = \exp\left(\frac{i}{2} \theta_{\mu\nu} \partial_{x_\mu}
\partial_{y_\nu}
\right)
f(x)g(y)\right\vert_{y=x}
\label{1}
\end{equation}
with
$\theta_{\mu\nu}$  a constant antisymmetric matrix.
Then, the Moyal bracket is defined as
\begin{equation}
\{f,g\} = f*g - g*f
\label{fg}
\end{equation}
The Moyal bracket for (Euclidean) space-time coordinates
then reads
\begin{equation}
\{x_\mu, x_\nu\} = i \theta_{\mu\nu}
\label{uno}
\end{equation}
In $4$-dimensional space one can always make
$\theta_{12} = \theta_1$,
$\theta_{34} = \theta_2$ (with $\theta_1$ and $\theta_2$ real numbers)
while all other  components vanish.

An alternative approach to noncommutative  theories which has
shown to be very useful in finding soliton solutions \cite{GMS}
is to directly work with operators in the phase space $\{x_\mu\}$
with commutator (\ref{uno}). Then, $*$ product is just
the product of operators and integration over $R^4$
becomes a trace,
\be
\int d^4x f(x) = 4\pi^2 \theta_1\theta_2  {\rm Tr} f(x)
\ee
In this framework, one
considers operators $a_b$ and $a_b^\dagger$ with $b=1,2$ in the form
\begin{eqnarray}
a_1 &=& \frac{1}{\sqrt {2\theta_1}} (x_1 + i x_2) = \frac{z_1}{\sqrt \theta_1}
\; , \;\;\;
a_1^\dagger = \frac{1}{\sqrt {2\theta_1}} (x_1 - i x_2) =
\frac{\bar z_1}{\sqrt \theta_1}\nonumber\\
a_2 &=& \frac{1}{\sqrt {2\theta_2}} (x_3 + i x_4) =
 \frac{z_2}{\sqrt \theta_2}  \; , \;\;\;
a_2^\dagger = \frac{1}{\sqrt {2\theta_2}} (x_3 - i x_4) =
 \frac{\bar z_2}{\sqrt \theta_2}
\label{as}
\end{eqnarray}
satisfying the algebra (consistent with (\ref{uno}))
\begin{eqnarray}
&& [a_b,a_c^\dagger] = \delta_{bc}  \;,  \;\;
[a_b,a_c] =0 \;, \;\; [a_b^\dagger,a_c^\dagger] = 0
\label{conmu}
\end{eqnarray}
With this conventions, derivatives should be written as
\begin{equation}
  \partial_{z_{b}}=\frac{-1}{\sqrt{\theta_{b}}}[a_{b}^{\dag}, ~]
 \; , \;\;\;
  \partial_{\bar{z_{b}}}=\frac{1}{\sqrt{\theta_{b}}}[a_{b}, ~]
  \label{deri}
\end{equation}
From the eigenstates of the number operators $N_1 = a_1^\dagger a_1$
and $N_2 = a_2^\dagger a_2$
one constructs the Fock space
$|n_1 n_2\rangle$ so that  $|00\rangle$ is
the  vacuum.  It will be important in what follows the following
correspondence between projectors and functions
\be
|n_1n_2 \rangle \langle n_1 n_2| \longrightarrow  4 (-1)^{n_1 + n_2}
\exp\left( - \frac{r_1^2}{\theta_1}  - \frac{r_2^2}{\theta_2}
\right)
L_{n_1}\left(\frac{2r_1^2}{\theta_1}\right)
L_{n_2}\left(\frac{2r_2^2}{\theta_2}\right)
\label{laguerre}
\ee
where $\vec r_1 = (x_1,x_2)$ and  $\vec r_2 = (x_3,x_4)$ and $L_n$ are
the Laguerre polynomials.
\vspace{0.5 cm}

Coming back to the instanton solution,
let us first see that a naive extension of the (commutative) $SU(2)$
't Hooft ansatz does not work in non-commutative space. To begin
with, it is well  known that consistency requires that the
gauge group for noncommutative
gauge theories has to be
$U(N)$ (or certain subgroups of $U(N)$ \cite{Bonora}-\cite{Jurco}).
We then consider the $U(2)$ case and write
\begin{equation}
A_\mu(x) = A_\mu^a \frac{\sigma^a}{2} + A_\mu^4 \frac{I}{2}
\end{equation}
The natural extension of the
commutative 't Hooft ansatz (\ref{tH})-(\ref{tHj}), to be
supplemented with an appropriate ansatz for
$A_\mu^4$,  is then
\begin{equation}
A_\mu^a \frac{\sigma^a}{2} =  \bar \Sigma_{\mu \nu}
J_\nu[\Phi]  \label{aa}
 \end{equation}
where
\begin{equation}
J_\nu  = \Phi^{-1} *\partial_\nu  \Phi  + \partial_\nu  \Phi * \Phi^{-1}
\label{Jd}
\end{equation}
Here we have taken a real $\Phi$ ($\Phi = \Phi^\dagger$) and hence
 the combination in (\ref{Jd}) leads to
  an Hermitian gauge field (In fact, (\ref{aa})-(\ref{Jd}) is
  the Hermitian version of the proposal in \cite{NS}).  Note that in the
  $\theta_1$, $\theta_2 \to 0$   limit, $J_\mu$ coincides with $j_\mu$
  in (\ref{tHj}) if one identifies $\Phi$ with $\phi^{1/2}$.

Concerning the $A_\mu^4$ choice, we shall use as a guide
that the appropriate ansatz  should lead,
together with $A_\mu^a$, to a self-dual $F_{\mu\nu}$ and in particular
a selfdual $F_{\mu\nu}^4$. Now, using
ansatz (\ref{aa})-(\ref{Jd}) one has
\begin{equation}
F_{\mu\nu}^4 =
\frac{i}{2}\{J_\mu ,J_\nu  \} +
\frac{i}{2}\varepsilon_{\mu\nu\alpha\beta}
\{J_\alpha ,J_\beta \} +
f_{\mu\nu}
\label{todo}
\end{equation}
where
\begin{equation}
f_{\mu \nu} = \partial_\mu A_\nu^4 - \partial_\nu A_\mu^4 + \frac{i}{2}
\{A_\mu^4,A_\nu^4\}
\label{fmu}
\end{equation}
Now, one can easily see that the simple ansatz
\begin{equation}
A_\mu^4 = -i \left(\Phi^{-1} * \partial_\mu \Phi - \partial_\mu \Phi *
 \Phi^{-1} \right)
\label{f2}
\end{equation}
leads to a self
dual $F_{\mu\nu}^4$  field,
\begin{equation}
F_{\mu\nu}^4 = \tilde F_{\mu \nu}^4 =
i\{J_\mu ,J_\nu \} +
\frac{i}{2}\varepsilon_{\mu\nu\alpha\beta}
\{J_\alpha ,J_\beta  \}
\label{sd}
\end{equation}
Concerning the  components of $F_{\mu\nu}$
on the Lie  algebra of $SU(2)$,
ansatz (\ref{aa})-(\ref{f2}) gives, in terms of $\Phi$,
\be
F_{\mu \nu}^a = \bar \eta_{a \mu \beta} V_{\beta \nu}
 -\bar \eta_{a \nu \beta} V_{\beta \mu}  -
 \frac{1}{2} \bar \eta_{a \mu\nu } V_{\beta \beta}
  -  \frac{1}{2} \bar \eta_{a\mu\nu} A
\label{efe}
\ee
where
\be
V_{\mu \nu} =
\left(\partial_\mu \partial_\nu \Phi^{-1}\right) * \Phi +
\Phi * \left(\partial_\mu\partial_\nu \Phi^{-1}\right) +
\partial_\mu \Phi * \Phi^{-2} * \partial_\nu \Phi +
\partial_\nu \Phi * \Phi^{-2} * \partial_\mu \Phi
\label{V}
\ee
and
\be
A =
 \Phi^{-1} * \nabla^2 \Phi^2 * \Phi^{-1}
\label{A}
\ee
Now, after some work, one can see that (\ref{efe})
can be written  in the form

\be
F_{\mu\nu}^a = \tilde F_{\mu\nu}^a - \bar \eta_{a\mu\nu} A
\label{nom}
\ee
so that, in order to satisfy selfduality, one just has to impose
\begin{equation}
  \Phi^{-1} * \nabla^2 \Phi^2 * \Phi^{-1} = 0
\label{sel2}
\end{equation}
This is the noncommutative version of the derivation summarized by
eqs.(\ref{vefe})-(\ref{fi}).
As in that case,  we conclude  now that if
one finds a field
$\Phi$ satisfying
\begin{equation}
\nabla^2 \Phi^2 = 0
\label{lapla}
\end{equation}
then  one has obtained an explicit solution for the $U(2)$
noncommutative instanton. Now, as explained for the
ordinary case, eq.(\ref{lapla}) has no nontrivial
solution so that one has to look for singular solutions (with
eventual sources in the r.h.s. of (\ref{lapla})) but still
satisfying (\ref{sel2}).

Paralleling the treatment in the ordinary case, one should then
introduce,
for finite $\theta_1$ and $\theta_2$, a regular, gaussian-like source
for the Laplacian, producing a delta-function  when
$\theta_1,\theta_2 \to 0$.
That is,  we propose to solve, instead of
a sourceless Laplace equation, the following one
\begin{equation}
\nabla^2\Phi^2(x;\theta_1,\theta_2) =
-\frac{4\lambda^2}{\theta_1\theta_2} \exp\left(-\frac{r_1^2}{\theta_1}
-\frac{r_2^2}{\theta_2}\right)
\label{antm}
\end{equation}
where $\lambda$ defines the instanton size.

When $\theta_1 = \theta_2=\theta$
one easily finds a solution to (\ref{antm}) in the
form
\begin{equation}
\Phi^2(x;\theta,\theta) = 1 + \frac{\lambda^2}{r_1^2 + r_2^2}\left(1-
\exp\left(-\frac{r_1^2 +r_2^2}{\theta}
\right)
\right)
\label{sol}
\end{equation}

Using eqs.(\ref{deri}) and (\ref{laguerre}) one can see that
 eq.(\ref{antm}) takes, in operator language, the form
\begin{equation}
\frac{2}{\theta_{1}}\left[a_{1}^{\dag},[a_{1},\Phi^2]\right]+
\frac{2}{\theta_{2}}\left[a_{2}^{\dag},[a_{2},\Phi^2]\right]=
\frac{\lambda^2}{\theta_1\theta_2}|00\rangle\langle
00|
\label{antmo}
\end{equation}
The right hand side
of eq.(\ref{antmo})  corresponds, in the Fock space framework,
to the Gaussian source introduced in eq.(\ref{antm}).

For $\theta_1 = \theta_2 = \theta$ the solution to (\ref{antmo})
can be written in a simple form
\begin{equation}
 \Phi^2(\theta,\theta) = 1 + \frac{\lambda^2}{2\theta} \sum_{n_1,n_2}
 \frac{1}{n_1 + n_2 + 1} |n_1 n_2 \rangle \langle n_1 n_2|
 \label{solF}
\end{equation}
Using sum rules for the Laguerre
polynomials associated to projectors, one can easily see that
(\ref{solF}) coincides with
 (\ref{sol}).
 Let us signal at this point that the $\theta_1 \ne \theta_2$ case
does not present new difficulties: one starts by solving eq.(\ref{antm})
and ends with the generalization of (\ref{solF}). We do not detail
this derivation since the results are conceptually equivalent to those
corresponding to $\theta_1 = \theta_2$.

Expression  (\ref{solF}) for $\Phi^2$ was originally presented in
\cite{NS} as providing an instanton solution
once the gauge field is written in terms of $\Phi$. Now, for
this to be true, one should verify that, although $\Phi$ does not
satisfy the sourceless equation (\ref{lapla}) but eq.(\ref{antm}),
it still verifies eq.(\ref{sel2}), which provides a necessary
condition for selfduality. Now, using expression (\ref{solF})
one finds
\be
\Phi^{-1} \nabla^2 \Phi^2 \Phi^{-1} =
-\frac{2\lambda^2}{\theta(2\theta +
\lambda^2)}
|00\rangle\langle 00|
\label{noda}
\ee
This implies that $F_{\mu \nu}$ is not selfdual but satisfies
\be
F_{\mu\nu} = \tilde F_{\mu \nu} + \bar \Sigma_{\mu \nu}
\frac{2\lambda^2}{\theta(2\theta +\lambda^2)}
|00\rangle\langle 00|
\label{joda}
\ee
A similar problem was found in \cite{LLY} in the investigation of
2 dimensional instantons in noncommutative $CP(n)$ model. Indeed, when
looking for a solution leading to a singular instanton
in the $\theta \to 0$ limit, these authors find that selfduality
was  satisfied up to a vacuum projector exactly as what happens with
the r.h.s. of eq.(\ref{joda}). In the ADHM approach to 4 dimensional
instantons, one also faces such
a problem but in that case, it manifests through a non-normalizable zero-mode.
However, in that case it is possible to find a ``shift'' transformation
which makes the zero mode normalizable \cite{F}-\cite{Furuu}.

Let us compute the topological charge associated with
the configuration (\ref{efe}),(\ref{solF}), using the formula
\be
Q = \frac{1}{4} \theta^2 {\rm Tr}\, {\rm tr}
 F_{\mu \nu} \tilde F_{\mu \nu}
\label{Qnc}
\ee
One finds, after a lengthy but straightforward calculation,
\be
Q = 0
\label{0}
\ee
It is interesting to note that $\lim_{\theta \to 0}\Phi^2 = \phi^{(1)}$
with $ \phi^{(1)}$ the solution leading to the singular
instanton solution (\ref{Asin}), which corresponds to $Q=1$. This shows
that it is not safe to interchange the
$\theta \to 0$ limit with the Tr operation.

At this point, one could think that projecting out from $F_{\mu \nu}$
its $|00\rangle\langle 00|$ one should obtain a selfdual curvature
with nontrivial ($n=1$) topological charge. However, if one just eliminates
the terms containing $|00\rangle\langle 00|$ from $F_{\mu \nu}$ one
obtains a selfdual expression but the corresponding Q is not integer
(and depends on $\lambda$ and $\theta$). This is due
to the fact  that such projected
$F_{\mu\nu}$ cannot be written as the curvature of an
adequate gauge connection. We haved also tried to find a kind of shift
 $S$ transformation, as defined in \cite{GN},\cite{AGMS},\cite{JKL},
 acting on $A_\mu$
 so that the resulting curvature has no $|00\rangle\langle 00|$
but we did not succeed in it.
In summary, in the form  proposed in \cite{NS}
 (eq.(4.9) in that paper)  or as modified   in (\ref{aa})-(\ref{f2}),
 the extension of 't Hooft ansatz  does not
 lead to a regular noncommutative instanton solution with $Q= 1$.

 In ordinary space, appart from 't Hooft ansatz, there is an alternative
 approach, at least for $Q=1$, which corresponds to search for
 a regular solution from the  begining, as done in the pioneering work of
 Belavin, Polyakov, Schwarz and Tyupkin \cite{BPST}.
 In the present case, this amounts to propose, instead of ansatz
 (\ref{aa}) one in the form
\begin{equation}
A_\mu^a \frac{\sigma^a}{2} =    \Sigma_{\mu \nu}
J_\nu[\Phi]  \label{aareg}
 \end{equation}
with $J_\mu$ defined as in (\ref{Jd}).
Note that, in contrast with ansatz (\ref{aa}) here we use
$\Sigma_{\mu \nu}$, the tensor arising in the regular
selfdual solution
in commutative space, eq.(\ref{Areg}). This ansatz leads to an
$F_{\mu\nu}^4$ in the form
\begin{equation}
F_{\mu\nu}^4 =
\frac{i}{2}\{J_\mu[\Phi] ,J_\nu[\Phi]  \} -
\frac{i}{2}\varepsilon_{\mu\nu\alpha\beta}
\{J_\alpha[\Phi] ,J_\beta[\Phi] \} +
f_{\mu\nu}[\Phi]
\label{todoreg}
\end{equation}
with $f_{\mu\nu}$ defined as in (\ref{fmu}).
Now, the choice for $A_\mu^4$ leading to a selfdual $F_{\mu\nu}^4$
is
\begin{equation}
A_\mu^4[\Phi] = i \left(\Phi^{-1} * \partial_\mu \Phi + 3\partial_\mu \Phi *
 \Phi^{-1} \right)
\label{f2reg}
\end{equation}
which is manifestly non-Hermitian. From ansatz
(\ref{aareg})-(\ref{f2reg}) one can construct $F_{\mu \nu}$ and
determine the conditions under which $F_{\mu\nu}^a$ is also selfdual.
One has
\be
F_{\mu\nu}^a[\Phi] = - \eta_{a\mu \nu} J_\alpha[\Phi] * J_\alpha[\Phi] +
\eta_{a\nu \alpha } D_{\mu \alpha}[\Phi] -
\eta_{a\mu \alpha } D_{\nu \alpha}[\Phi]
\label{ese}
\ee
where
\be
D_{\mu \alpha} = -\Phi * \partial_\mu \partial_\alpha \Phi^{-2} * \Phi
+ \{J_\mu[\Phi],J_\alpha[\Phi]\}
\label{da}
\ee
Now, it is easy to see from the expression for
$F_{\mu\nu}^a$ as given by (\ref{ese})  that selfduality is ensured
whenever the symmetric part of $D_{\mu\alpha}$ satisfies
\be
-\Phi * \partial_\mu \partial_\alpha \Phi^{-2} * \Phi = D[\Phi]
\delta_{\mu\alpha}
\label{sim}
\ee
where $D[\Phi]$ is an arbitrary function. Equation (\ref{sim})
has the simple solution
\be
\Phi^{-2} = 1 + \frac{1}{\lambda^2} \left(r_1^2 + r_2^2\right)
\label{star}
\ee
which in the operator language reads
\be
\Phi^{-2} = 1 + \frac{2\theta}{\lambda^2}\sum_{n_1 n_2} \left(
n_1 +n_2 +1\right)|n_1 n_2 \rangle\langle n_1 n_2|
\label{op}
\ee
With this expression one can compute $A_\mu$ from eq.(\ref{aareg})
and then the field strength which reads
\be
F_{\mu\nu}^a = \eta_{a\mu\alpha } \{J_\nu,J_\alpha\} -
 \eta_{a\nu\alpha} \{J_\mu,J_\alpha\} - \eta_{a\mu\nu} \left(
 J_\alpha *J_\alpha + 2 D[\Phi]
 \right)
\label{FnoH}
\ee
where $D$, as defined by (\ref{sim}) takes for the solution (\ref{star})
the form
\be
D = -\frac{2}{\lambda^2} \Phi^2
\label{DDD}
\ee

Now, as one should expect from (\ref{f2reg}), $F_{\mu\nu}$ is not
Hermitian. However the Lagrangian and, a fortiori, the action $S$ and
the topological charge $Q$, are real. In fact, one can check by explicit
computation that $S=Q=1$, a result consistent with the fact that
$F_{\mu\nu}$ as given by (\ref{FnoH}) is a selfdual curvature which,
necessarily,
satisfies the (noncommutative) Yang-Mills equations of motion.

We have discussed
the case $\theta_1 = \theta_2$ for
the sake of simplicity, but the general $\theta_1 \ne \theta_2$ case can be
equally treated just by noting that the solution (\ref{op})  becomes
\be
\Phi^{-2} = 1 + \frac{1}{\lambda^2}\sum_{n_1 n_2} \left(
2\theta_1 n_1 +
2\theta_2 n_2 +\theta_1+\theta_2 \right)
|n_1 n_2 \rangle\langle n_1 n_2|
\label{opdif}
\ee

We have seen that  noncommutative versions of 't Hooft ansatz yields to
$U(2)$ configurations which are either non selfdual or non Hermitian. One
might then expect that  a less restrictive ansatz could overcome these problems.
One possibility is to
consider, instead of (\ref{aareg})
\begin{eqnarray}
A_\mu &=& \Sigma_{\mu\nu} a_\nu\nonumber\\
A_\mu^4 &=& b_\mu
\end{eqnarray}
with $a_\mu$ and $b_\mu$ Hermitian. The curvature
$F_{\mu\nu}^4$ takes then the form,
\begin{equation}
F_{\mu\nu}^4 =
\frac{i}{2}\{a_\mu ,a_\nu  \} -
\frac{i}{2}\varepsilon_{\mu\nu\alpha\beta}
\{a_\alpha ,a_\beta \} +
\partial_\mu b_\nu - \partial_\nu b_\mu + \frac{i}{2}\{b_\mu,b_\nu\}
\label{todoab}
\end{equation}
Now, the selfduality condition applied to $F_{\mu\nu}^4$
implies a relation
between $a_\mu$ and $b_\mu$ as well as a condition over their curvatures.
A simple
choice satisfying all these restrictions is
\be
b_\mu = a_\mu
\ee
\be
\partial_\mu a_\nu - \partial_\nu a_\mu + 2i \{a_\mu,a_\nu\} = 0
\label{esto}
\ee
Now, in order to also achieve selfduality for $F_{\mu\nu}^a$, the following
identity should hold (compare with (\ref{da})-(\ref{sim}))
\be
\partial_\mu a_\nu + \partial_\nu a_\mu - \{a_\mu,a_\nu\}_+ =
D[a] \delta_{\mu\nu}
\label{lootro}
\ee
Concerning condition (\ref{esto}), it is trivially satisfied by
\be
a_\mu = - \frac{i}{2} U(x)^\dagger * \partial_\mu U(x)
\label{gauge}
\ee
with $U(x)$ an element of noncommutative $U(1)$ group. In terms of
$U(x)$, eq.(\ref{lootro}) becomes
\be
-\frac{i}{2} U^{-1} \partial_\mu \partial_\nu U -
\frac{1}{8} ({1} + 2i) \left(\partial_\mu U^{-1} \partial_\nu U
+\partial_\nu U^{-1} \partial_\mu U\right) = D[U] \delta_{\mu \nu}
\label{hoe}
\ee
Unlike equation (\ref{sim}), we were unable to find a solution of eq.
(\ref{hoe}).

In summary, we have shown that the natural extension of the
't Hooft ansatz for  $U(2)$ instanton solutions to noncommutative  spacetime,
as proposed in \cite{NS}, does not work since it leads to a non-selfdual (or
self-antidual) field strength. An alternative ansatz allows
to find  a selfdual
instanton solution which however corresponds to a non-Hermitian gauge field.
Nevertheless, this configuration leads to a real Lagrangian and corresponds
to a bound of the action S: its topological charge is $Q=S=1$.

The connection between commutative and noncommutative instantons can
then be schematized as follows:
the extension of 't Hooft ansatz  leading to singular instantons yields, in the
noncommutative case, to non-selfdual configurations. Concerning the ansatz
leading to regular ordinary instantons, it  does give,
in the noncommutative case, selfdual solutions.
The explicit one that we found was  not Hermitian but one should expect
that more general ansatz would lead to selfdual Hermitian instantons.
We hope to come back to this problem in the future.

\vspace{1 cm}

\noindent\underline{Acknowledgements}:
 We acknowledge N.~Nekrasov for usefull correspondence. This work  partially
 supported
 by UNLP, CICBA, CONICET (PIP 4330/96), ANPCYT (PICT 03-05179), Argentina.
 G.S.L. and E.F.M. are partially supported by Fundaci\'on Antorchas, Argentina.

\newpage

%%%%%%%%%%%%%%%%%%%%%%%%%%%%%%%%%%%%%%%%%%%%%%%%%%%%%%%%%%%%%%%%%

\end{document}